\documentclass[a4paper]{jpconf}
\usepackage{graphicx}
\usepackage{nicefrac}
\usepackage{enumitem}

\begin{document}

\title{Superfluid density in cuprates: hints on gauge compositeness of the holes}
\author{P A Marchetti and G Bighin}
\address{Dipartimento di Fisica e Astronomia, Universit\`a di Padova 
 and INFN, I-35131 Padova, Italy}
\ead{marchetti@pd.infn.it, bighin@pd.infn.it}

\begin{abstract}
We show that several features (the three-dimensional XY universality for moderate underdoping, the almost-BCS behaviour for moderate overdoping and the critical exponent) of the superfluid density in hole-doped cuprates hint at a composite structure of the holes. This idea can be implemented in a spin-charge gauge approach to the $t-t'-J$ model and provides indeed good agreement with available experimental data.
 \end{abstract}

\section{Superfluid density: a puzzle and a solution}

In this work we point out some features of the in-plane superfluid density $\rho_s$ in hole-doped high-$T_c$ cuprates  which hint at, or at least are fully compatible with, a peculiar ``gauge compositeness" of
the low-energy hole excitations in such materials.
A similar suggestion comes also from transport and entropy arguments as discussed in Ref. \cite{masy}.

\subsection{Superfluid density in underdoped cuprates: theoretical constraints from experiments.}
In the region from moderate underdoping to optimal doping the superfluid density as a function of the temperature $\rho_s(T)$  exhibits a linear $T$-dependence near $T=0$, along with the critical exponent $\nicefrac{2}{3}$ at the critical temperature $T_c$ \cite{kamal}. In the same doping region, the normalized superfluid density $\rho_s(T/T_c)/\rho_s(0)$  shows a non-BCS, 3DXY-like universality independent both of doping concentration and of the specific kind of material involved \cite{hardy,mb}. Finally, the Uemura linear relation \cite{uemura} between $\rho_s(T=0)$ and $T_c$ approximately holds in underdoped cuprates.

These features put severe constraints on theoretical explanations of the behaviour of $\rho_s$. Specifically a BCS-based explanation of the $T$-linear dependence at low temperatures, as due to the quasi-particle excitations near the nodes of the $d$-wave BCS order parameter \cite{lee}, is difficult to reconcile with the non mean-field critical exponent (which should be $1$ according to the BCS theory), with the observed universality and with the Uemura relation, as within the BCS theory $\rho_s(0)$ does not depend on the order parameter controlling $T_c$.

An alternative explanation of the $T$-linear behaviour is based on phase or pairing fluctuations \cite{kivel} of the order parameter in a BCS-BEC crossover setting, resulting in an effective low-energy XY model. However the most natural XY model obtained is two-dimensional, thus producing an incorrect critical behaviour. A three-dimensional nature of the XY model is sometimes claimed to emerge in a narrow range of temperatures close to $T_c$ due to the presence of a stack of $\mathrm{Cu}$-$\mathrm{O}$ layers, but this is not sufficient to explain the three-dimensional XY universality of the normalized $\rho_s$ over the entire temperature range from $T=0$ to $T_c$. Furthermore, a well defined gapped Fermi surface seen in ARPES experiments \cite{damascelli} suggests that cuprates should lie quite close to the weakly-coupled region of the BCS-BEC crossover and seems hard to reconcile with the observed large pseudogap region above $T_c$.
We also stress that in BCS-BEC approaches, the translation and time-reversal symmetries being unbroken, a theorem due to Leggett \cite{leggett} should apply, stating that the superfluid fraction at $T=0$ is unity. However this is at odds with the Uemura relation even if we assume the charge carriers to be given only by the doped holes, as suggested by transport measurements in the underdoping region \cite{taka}.

For slightly overdoped samples the above quoted universality disappears and the normalized superfluid density is BCS $d$-wave-like over a broad temperature range. Nonetheless the critical exponent at $T_c$ appears still to be $\nicefrac{2}{3}$ and the critical temperature obtained extrapolating the low-temperature behaviour to higher temperatures assuming fully BCS behaviour is always larger than the actual one \cite{stajic}, approaching the BCS value only at sufficiently high overdoping.

\subsection{A solution: spin-charge gauge approach to superfluid density.}
We now describe in a nutshell how a possible explanation of the above puzzling features is provided by a composite structure of the holes, physically identified with Zhang-Rice singlets, arising in a spin-charge gauge approach \cite{jcmp,mysy,mg} to the $t-t'-J$ model for the $\mathrm{Cu}\mathrm{O}_2$ planes.  In such framework the holes are described in the low-energy limit as bound states of a charge excitation, the holon, and spin excitation, the spinon. The holon carries the essential informations about the Fermi surface while the gapped spinon carries the spin degrees of freedom and determines the character of the critical transition. The binding force between holon and spinon is due to a slave-particle gauge field arising from the no-double occupation constraint emerging from a $t-J$ model description of Zhang-Rice singlets.
In the superconducting phase both holons and spinons are paired. The dynamics of the phase of holon pairs is BCS-like whereas the dynamics of the phase of the spinon pairs is described by a three-dimensional (gauged) XY model whose inverse coupling is a function $\Theta(T)$ of the physical temperature $T$, of the spinon gap and the spinon-pair density. The critical temperature $T_c$ is found imposing the criticality condition for the XY model, thus being determined by the spinonic part of the theory.

Within this formalism the superfluid density has a contribution both from holon-pairs, $\rho_s^h$, and from spinon-pairs, $\rho_s^s$; due to the gauge ``string'' binding  holons to spinons, the two contributions add in parallel, obeying Ioffe-Larkin \cite{ioffe} rule
\begin{equation}
\rho_s =  \frac{\rho_s^h \rho_s^s}{\rho_s^h + \rho_s^s}
\label{eq:il}
\end{equation}
this feature being typical of spin-charge approaches, as also derived e.g. in Ref. \cite{lee2}.

In the moderate underdoping region the spinon contribution to the superfluid density turns out to be dominant \cite{mb}. Gauge fluctuations are gapped by the Anderson-Higgs mechanism and it turns out that $\Theta(T)/\Theta(T_c) \approx T/T_c$. This last property is critical in determining that the normalized superfluid density profile is 3DXY-like across the whole temperature range from $T=0$ up to the critical temperature $T_c$. The spinons are insensitive to the Fermi surface details, recorded by holons, explaining their universality. Both the linear $T$-dependence near $T=0$ and the critical exponent $\nicefrac{2}{3}$, characteristic of the three-dimensional XY model, then follow naturally. There is no problem for coexistence of the Fermi surface, due to holons, with a large pseudogap since the $T_c$ scale is set by spinon-pair condensation while the BCS-BEC-like pseudogap is due to holon pairing. Furthermore the XY contribution to the superfluid density at $T=0$ is of the form $(\mathrm{d} \Theta/ \mathrm{d} T(0))^{-1}$ as argued in Ref. \cite{mb}, while $T_c$ is determined by the XY transition for the effective temperature $\Theta(T)$. Therefore if we denote by $T_c^{XY}$ the critical temperature of the three-dimensional XY model, we have $T_c^{XY}=\Theta(T_c) \approx (\mathrm{d} \Theta / \mathrm{d} T(0)) \ T_c$, and an approximate Uemura relation follows \cite{mb}.

In moderately overdoped samples the holon contribution becomes dominant, except close to $T_c$, we thus recover the more standard $d$-wave BCS structure for the normalized superfluid density. Universality is lost due to the sensitivity of holons to Fermi surface details. The scale of the superconducting transition is still set by spinon pair condensation, whence the $\nicefrac{2}{3}$ critical exponent. However the pseudogap temperature, setting the scale for holon pairing, is larger than $T_c$ and this explains why the critical temperature obtained extrapolating from the BCS formula is always larger than the real $T_c$.
In Fig. \ref{fig:1} we show experimental data for the normalized superfluid density for several different samples, compared with the three-dimensional XY model and the two-dimensional $d$-wave BCS theory.

\begin{figure}[ht]
  \begin{minipage}[c]{0.57\textwidth}
    \includegraphics[width=\textwidth]{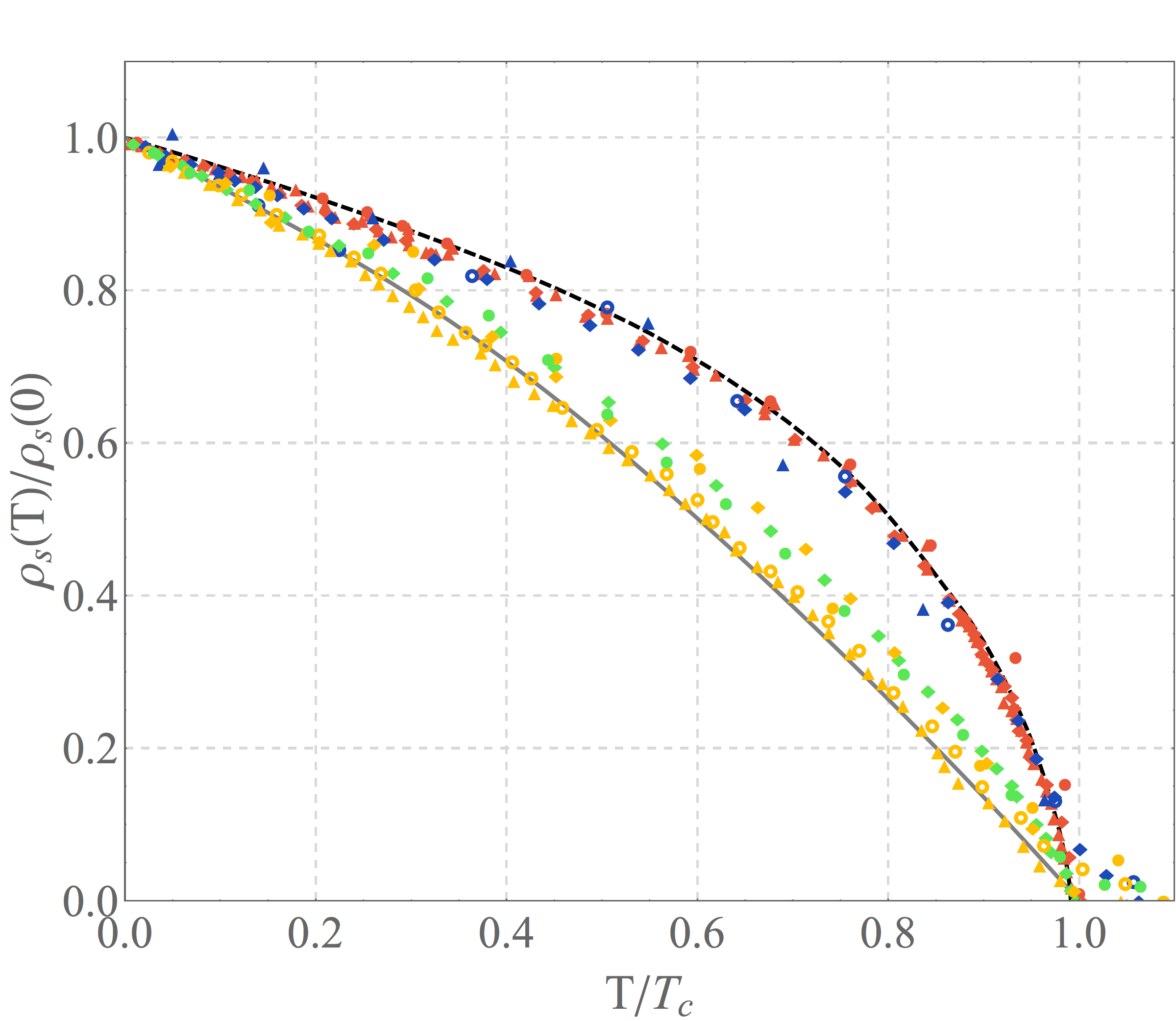}
  \end{minipage}\hfill
  \begin{minipage}[c]{0.40\textwidth}
\caption{\footnotesize \label{fig:1} Superfluid density: experimental data for several samples compared with the 3DXY model (black dashed line) and the two-dimensional BCS $d$-wave theory (gray line). Red markers denote underdoped YBCO samples, precisely $x=6.95$ $a$-axis from \cite{zhang} (diamonds), underdoped $a$-axis (filled circles) and universal $a$-axis behaviour (triangles) from \cite{hardy}; blue markers denote other under- or optimally-doped samples, namely Bi-2112 from \cite{jacobs} (diamonds), LSCO $x=0.15$ from \cite{lemberger} (filled circles), Hg-1201 $x=0.10$ (triangles) and LSCO $x=0.15$ from $\mu$SR (open circles) from \cite{panagopoulos}. Yellow markers denote overdoped samples from \cite{panagopoulos}, namely Hg-1201 $x=0.154$ (diamonds), Hg-1201 $x=0.37$ (filled circles), LSCO $x=0.20$ (triangles), LSCO $x=0.22$ (open circles). Finally green markers denote overdoped YBCO, $x=6.99$, $b$-axis from \cite{hardy} (diamonds) and Y124 from \cite{shengelaya} (circles).} \label{fig:1}
  \end{minipage}
\end{figure}

\section{Spin-charge gauge approach to superconductivity}

Let us briefly summarize the key steps of this approach to superconductivity, giving only few details useful to get a clue for the statements made above; for an alternative brief presentation of the ideas involved the reader is referred to Ref. \cite{pam}:
\begin{enumerate}[topsep=0pt,itemsep=-1ex,partopsep=1ex,parsep=1ex,leftmargin=0pt,itemindent=3em]
\item Spin-charge decomposition. To satisfy the constraint of no-double occupation in the $t-t'-J$ model, we use the slave-particle formalism splitting the fermionic hole field $c$ into a product of a fermionic spinless charged field, the holon $h$, and a bosonic neutral spin $\nicefrac{1}{2}$ field, the spinon $s$
\begin{equation}
c_{j, \alpha} = s_{j, \alpha}^* h_j
 \; ,
\end{equation}
bound together in two dimensions by gauge interactions mediated by a slave-particle gauge field $A_\mu$. The Euclidean effective action of the spinons in the continuum limit is given by an O(3) $\sigma$ model, with ``relativistic'' dispersion treating space and time on the same footing.

\item Semionic statistics. In one dimension the correct statistics for both fields is semionic \cite{hmsy}, exactly intermediate between bosonic and fermionic. The realization of this statistics in two dimensions 
is obtained binding a statistical $\nicefrac{1}{2}$ charge flux $\Phi^h$ to the holons corresponding to empty sites and a statistical $\nicefrac{1}{2}$ spin flux $\Phi^s$ to the spinons
\begin{equation}
c_{j \alpha} = e^{\mathrm{i} \Phi_h (j)} h_j  \left( e^{- \mathrm{i} \Phi_s (j)} s_j^* \right)_\alpha \; ,
\end{equation}
still retaining the fermionic statistics for the holes.  
These semionic holons have recently been proved \cite{fmsy} to obey Haldane statistics of order 2 in momentum space, meaning that a maximum of two semions are permitted to have the same momenta. Hence a gas of spinless semions of finite density has a Fermi surface at low $T$ coinciding with that of spin $\nicefrac{1}{2}$ fermions with the same density, thus recovering the tight binding Fermi surface often used in the discussion of the experimental data, if the $J$ coupling is neglected. In the approximate treatment developed in Ref. \cite{jcmp}, followed here, after taking into account this exclusion effect any semionic character of holons and spinons is neglected. 

\item Mean-field treatment. 
In the adopted mean-field approximation (MFA) we neglect the
holon fluctuations in $\Phi^h$ and the spinon fluctuations in
$\Phi^s$ . Then $\Phi^h$ is static and provides a $\pi$-flux phase factor per
plaquette. This flux yields for the holons 
two small Fermi surfaces, $\epsilon_F \sim
t\delta$, centered at $(\pm \pi/2, \pi/2)$, characterizing the ``pseudogap phase'' (PG) of the model. Increasing doping or temperature one reaches a crossover line $T^*$, identified with the experimental inflection point of in-plane resistivity \cite{jcmp}, 
above which we enter the ``strange metal phase'' (SM) where  the effect of the charge flux is screened by spinons and a ``large'' tight-binding Fermi surface for the holons is recovered, with $\epsilon_F \sim t (1+\delta)$.
 The spin flux in MFA is given by:
\begin{eqnarray}
\label{eq:5}
\Phi^s(x) = \sigma_z \sum_l
{h}^{*}_l {h}_l\frac{(-1)^{|l|}}{2} 
\arg(\vec{x}-\vec{l}),
\end{eqnarray}
hence it attaches to the empty lattice sites quantum spin vortices with opposite vorticity if centered on holons in different N\'eel sublattices. These vortices appear in the U(1) subgroup of the spin group complementary to the coset  labeling the directions of the spin.

\item Short range anti-ferromagnetic (AF) order and charge pairing. The interaction term between spinons and spin-vortices in the continuum limit is of the form 
\begin{eqnarray}
\label{spvo}
J (1-2 \delta)(\nabla{\Phi^s(x)})^2  s^* s \; .
\end{eqnarray}
Averaging the spin-flux contribution of this term one obtains a mass gap, $ m_s^2 \approx 0.5 \ \delta |\log \delta|$, for  the spinons reproducing the short-range AF order caused by doping.   By averaging instead the spinons in Eq. (\ref{spvo}), we obtain an effective interaction:
\begin{equation}
\label{zh}
J (1-2 \delta) \langle s^* s \rangle \sum_{i,j} (-1)^{|i|+|j|} \Delta^{-1}
 (i - j) h^*_ih_i h^*_jh_j,
\end{equation}
where $\Delta$ is the two-dimensional lattice Laplacian, which yields a $d$-wave pairing between the charges associated with spin vortices centered on different N\'eel sublattices.  This charge pairing occurs below a temperature $T_{ph}$, well comparing with the experimental (upper) pseudogap temperature,  and it reproduces the phenomenology of  Fermi arcs coexisting with a pseudogap in the antinodal region \cite{mg} which adds to the $\pi$ flux pseudogap in PG.
The origin of the charge-pairing is magnetic, but it is not due to exchange of AF spin fluctuations.
  
\item Spin pairing.
The spins of the charge pairs turn into local RVB spin-singlets,  only at a lower temperature $T_{ps}<T_{ph}$, well comparing with the experimental onset of Nernst signal \cite{mysy}, as effect of a binding force between holons and spinons.
The lowering of
free energy allowing the formation of spinon pairs is due to the appearence of short-range vortex-antivortex pairs which do not contribute to the spinon gap in Eq. (\ref{spvo}). 

\item Superconductivity.  It occurs by condensation of hole (i.e. holon+spinon) pairs at a temperature $T_c<T_{ps}$. Below $T_{ps}$, since spinons are gapped with ``relativistic'' dispersion and spinon pairs induce an RVB order parameter $\Delta_s$. The Euclidean effective Lagrangian obtained integrating out the spinons is then a gauged  three-dimensional XY model of the form
\begin{equation}
\frac{1}{6 \pi M_s} \{ [ \partial_\mu A_\nu - \partial_\nu A_\mu ]^2 + | \Delta_s |^2 [ 2 ( A_0 + \partial_0 \frac{\phi}{2} )^2 + ( \mathbf{A} + \nabla \frac{\phi}{2} )^2 ] \} \; ,
\end{equation}
where $\phi$ is the condensate phase and $M_s \approx m_s-| \Delta_s |^2/m_s$.
Since in the superconding phase $A$ is gapped, it follows that superconductivity appears when $\Theta (T)=(| \Delta_s |^2 / 3 \pi M_s)^{-1}\approx T_c^{XY}$. At $T=0$, interestingly, $| \Delta_s | \approx m_s$, so that $M_s \approx 0$, suggesting that all spin vortices, and hence spinons, are paired, and the aforementioned theorem due to Leggett is obeyed. The theorem is also satisfied independently by holons, because of their BCS-like nature. However it is not true that the superfluid number density equals the hole density, due to the Ioffe-Larkin addition rule for the holon and the spinon contributions to $\rho_s$ in Eq. (\ref{eq:il}).

The spinonic contribution to superfluid density entering the Ioffe-Larkin rule in Eq. (\ref{eq:il}) is $\rho_s^s \sim (\mathrm{d} | \Delta^s |^2/\mathrm{d} T)^{-1}$. In the PG regime $| \Delta^s |^2$ is essentially linear in $T$, implying that the leading contribution to $\rho_s$ comes from spinons, while on the other hand in SM $| \Delta^s |^2$ is essentially constant in $T$ almost to $T_c$, implying that the leading contribution to $\rho_s$ comes from holons, except near $T_c$. This explains the XY-like behaviour in underdoped samples and the BCS-like behaviour in overdoped samples.
\end{enumerate}

Summarizing in this approach the hole is a composite of a  ``relativistic'' gapped bosonic spin $\nicefrac{1}{2}$ spinon and a spinless charge holon with a Fermi surface. This compositeness is reflected in the superconductivity mechanism: lowering $T$ first charge-pairing occurs, followed by spin-pairing; at an even lower temperature $T_c$ spin-pair condensation finally occurs, leading to superconductivity. The compositeness is reflected also in the superfluid density and using spinon dominance in the underdoping region (PG) and holon dominance in the slightly overdoping region (SM) we are able to simultaneously explain many several experimental features of $\rho_s$. 

\section*{Acknowledgements}
PAM thanks Z.~B.~Su, L.~Yu  and Y.~Fei for the joy of a long collaboration on this project and I. Bo\v{z}ovi\'c for an illuminating discussion. Authors acknowledge partial support from PRIN Project ``Collective Quantum Phenomena: from Strongly-Correlated Systems to Quantum Simulators''.

\end{document}